\documentclass[pra,twocolumn,floatfix]{revtex4}
                        \usepackage{graphicx}
                        \usepackage{dcolumn}
                        
\begin{document}

                        \def\be{\begin{equation}}
                        \def\ee{\end{equation}}
                        \def\ba{\begin{eqnarray}}
                        \def\ea{\end{eqnarray}}
                        \def\bas{\begin{eqnarray*}}
                        \def\eas{\end{eqnarray*}}


\title{The Wavefunction of the Collapsing Bose-Einstein Condensate}

\author{Stavros Theodorakis and Andreas Hadjigeorgiou}
                        \affiliation{Physics Department, University of Cyprus,
P.O. Box 20537, Nicosia 1678, Cyprus}
                        \email{stavrost@ucy.ac.cy}
\date{\today}

\begin{abstract}
Bose-Einstein condensates with tunable interatomic interactions have been studied intensely in recent experiments. The investigation of the collapse of a condensate following a sudden change in the nature of the interaction from repulsive to attractive has led to the observation of a remnant condensate that did not undergo further collapse. We suggest that this high-density remnant is in fact the absolute minimum of the energy, if the attractive atomic interactions are nonlocal, and is therefore inherently stable. We show that a variational trial function consisting of a superposition of two distinct gaussians is an accurate representation of the wavefunction of the ground state of the conventional local Gross-Pitaevskii field equation for an attractive condensate and gives correctly the points of emergence of instability. We then use such a superposition of two gaussians as a variational trial function in order to calculate the minima of the energy when it includes a nonlocal interaction term. We use experimental data in  order to study the long range of the nonlocal interaction, showing that they agree very well with a dimensionally derived expression for this range.
\end{abstract}

\maketitle

\vskip 0.3cm
\vskip 0.3cm

{\bf I. Introduction}

Attraction between the atoms of a Bose-Einstein condensate renders it unstable, although a condensate with a
limited number of atoms can be stabilized by confinement in an atom trap. However, beyond a critical number of atoms the condensate
collapses. This collapse has been investigated extensively in recent years\cite{ThompsonDonley}, in the context of Bose-Einstein condensates with tunable interatomic interactions. In the vicinity of a Feshbach resonance, the scattering length depends sensitively on the magnitude of an externally applied magnetic field, allowing the magnitude and sign of the atomic interactions to be tuned to any value. Such a resonance in $^{85}$Rb has been exploited in order to investigate the collapse of a condensate following a sudden change in the nature of the interaction from repulsive to attractive.

Typically in such a process the initial scattering length $a_{init}$ is $0$, with the Bose-Einstein condensate (BEC) taking on the size and shape of the harmonic oscillator ground state. It is suddenly changed to a negative scattering length $a_{collapse}$. This sudden change leads to a violent collapse process, in which a significant proportion of the initial condensate is ejected in a highly energetic burst during the early stages of the collapse, leaving behind though a high-density remnant condensate that survives for many seconds. It was not understood why this remnant BEC did not undergo further collapse. The number of atoms in the remnant was observed not to be limited by any critical number. In fact, a fixed fraction of the initial number of atoms $N_{0}$ went into the remnant, independent of $N_{0}$. This fraction decreased with $| a_{collapse} |$, and was about 40$\%$ for $| a_{collapse} |<10 a_{0}$ and about $10\%$ for $| a_{collapse} | > 100 a_{0}$, where $a_{0}$ is the Bohr radius \cite{nature}.

Calculations of the loss of atoms during the collapse have been made \cite{Adhikari} by including an absorptive nonlinear three-body recombination term in the local Gross-Pitaevskii (GP) equation. The atom loss is simulated by an quintic term in the field equation with an imaginary coefficient. The number of the atoms in the remnant can indeed be much larger than the critical number in these calculations, in accordance with the experimental observations. However, in these simulations the remnant continues to emit atoms, and for very large times on the order of seconds the number of atoms eventually tends toward the critical number. In contrast, the observed remnant survives the collapse and appears to be pretty stable.

An alternative explanation for the unexpected stability of remnant condensates for which the remnant number exceeds the critical number is that the remnant condensate is composed of multiple solitons that repel each other\cite{Thompson}. The number of atoms in each soliton never exceeds the critical number and the repulsive solitons never overlap, thus the condensate stability condition is never violated. Indeed, the remnant condensate was experimentally observed to separate into two or more distinct clouds which were considered to be associated with solitons. Neighboring solitons supposedly form with a relative phase that ensures that they interact repulsively, even though the atomic interactions are attractive. Consequently, as the solitons never fully overlap, the critical density for collapse is never reached and the individual solitons remain stable, the number of atoms in each soliton being always less than or equal to the critical number. Theoretically, the two mutually repelling solitons have been represented by a first excited state, since the density profile of such a state is double-peaked and can be interpreted as two solitons featuring a $\pi$-phase difference\cite{Parker}. However, there is no definite physical picture of the origin of this relative phase that is needed to ensure the stability of the multiple soliton states.  In fact, the dynamical creation of solitons does not favour a repulsive relative phase\cite{Dabrowska}, because the even symmetry of the initial GP wavefunction for this particular experiment\cite{Thompson} prevents repulsive phase relations in the final state for an even number of solitons, as mean-field theory preserves this symmetry. In this situation therefore, the central two solitons must have zero relative phase. Furthermore, a recent analysis\cite{Gardiner} revealed that the anisotropy of the particular experiment is too small to achieve a highly solitonlike ground state.  According to this analysis, the regime in which the ground state is highly solitonlike and amenable to a macroscopic superposition of solitons is significantly restricted and occurs only for experimentally challenging trap anisotropies.

In this paper we propose an alternative explanation for the remarkable stability of the large remnant condensates. We show that if the GP equation is modified so as to include nonlocal interactions, then there is no collapse, but just a transition to a high-density absolute minimum of the nonlocal energy. Since the change is sudden, the fraction of atoms remaining in the remnant condensate is simply the square of the modulus of the overlap integral between the initial harmonic oscillator state and the final state that corresponds to the absolute minimum of the nonlocal energy.

In order to compare the predictions from this idea with the experimental data, we need an accurate wavefunction for the collapsing condensate. We show thus first that a superposition of two distinct gaussians is an excellent variational trial function for the final wavefunction of a collapsing BEC that is described by the local anisotropic GP equation. We then use such a superposition of two distinct Gaussians, with entirely different parameters, as a variational trial function for the final wavefunction when the energy includes a nonlocal interaction term. This final wavefunction that consists of two gaussian components will evolve, of course, after the collapse. Thus the coefficients of the two gaussian components will change with time, possibly leading to configurations where they will have different signs, in which case the wavefunction will have a node. The corresponding density will then have two peaks separated by a time-dependent domain of zero values, giving thus the impression of two separate oscillating solitons.

In section II, we present the nonlocal energy that describes the condensate. We show in section III that when the interaction is strictly local a superposition of two distinct gaussians is an accurate representation of the exact wavefunction of the attractive condensate, both for isotropic and anisotropic trap potentials. In section IV we extend our work to the nonlocal case, calculating the energy when the interaction is nonlocal. In section V we examine the special case of a single Gaussian trial function, in order to understand qualitatively the main physics introduced by the nonlocal interaction. In section VI we show that while the single Gaussian cannot agree with the experimental data on remnant condensates, the superposition of two Gaussians can. In fact, we use these data in order to find the dependence of the range of nonlocal interactions on the scattering length. We derive from dimensional arguments an expression relating the range of interactions to the scattering length and we show that it agrees very well with the experimental data. Section VII summarizes our conclusions.

\vskip 0.3cm
\vskip 0.3cm

{II. \bf The nonlocal energy}

Nonlocality in the effective interaction is a crucial ingredient that prevents collapse when the scattering length is negative. Long-range potentials favor the formation of big clouds that still maintain a rather low density. When three extrema of the full energy are present, the intermediate one represents an unstable state (a local maximum of the energy), while the other two respectively describe a low-density metastable solution (local minimum) and a higher-density stable solution (absolute minimum)\cite{Salasnich}. Thus a nonlocal attractively interacting condensate cannot collapse\cite{Konotop}, even in the absence of a trap potential\cite{Turitsyn}. At worst, as we shall demonstrate in Section V, a transition may occur between the two classes of stable solutions.

Let us begin with the energy functional for an nonlocally attractive Bose-Einstein condensate in an anisotropic harmonic trap:
            \ba
            \label{E3D}
           &&E=\int\,\,d^{3}r\,\big(\frac{\hbar^{2}}{2m}|\nabla\Psi|^{2}+\frac{1}{2}m(\omega_{r}^{2}{\bf r_{\perp}}^{2}+\omega_{z}^{2}z^{2})|\Psi|^{2}\nonumber\big)\\
           &&-\frac{g}{2}\int\,\,d^{3}r\,\int\,\,d^{3}r^{\prime}\,|\Psi({\bf r})|^{2} V(|{\bf r}-{\bf r^{\prime}}|)|\Psi({\bf r^{\prime}})|^{2}
            \ea
            where $\int\,d^{3}r|\Psi|^{2}=N_{0}$, $N_{0}$ being the fixed number of particles and ${\bf r_{\perp}}$ being the position vector in the xy plane. If $a$ is the negative scattering length, then $g=4\pi |a|\hbar^{2}/m$. We set $\Psi=\sqrt{N_{0}}\psi({\bf r})/\sqrt{d^{3}}$, where $d=\sqrt{\hbar/(m\omega)}$, with $\omega=\omega_{r}^{2/3}\omega_{z}^{1/3}$. For the sake of convenience we render the various variables dimensionless. Thus we measure the distances in units of $d$, the interaction potentials in units of $1/d^{3}$ and the energies in units of $N_{0}\hbar\omega$. We define the dimensionless parameters $k=|a|N_{0}/d$ and $\lambda=\omega_{z}/\omega_{r}$, noting that $\psi$ is also dimensionless. Then $\int\,d^{3}r|\psi|^{2}=1$ and the energy reduces to the dimensionless form
						
						 \ba
            \label{E3Ddimless}
           &&E=\int\,\,d^{3}r\,\big(\frac{1}{2}|\nabla\psi|^{2}+\frac{1}{2}(\lambda^{-2/3}{\bf r_{\perp}}^{2}+\lambda^{4/3}z^{2})|\psi|^{2}\nonumber\big)\\
           &&-2\pi k\int\,\,d^{3}r\,\int\,\,d^{3}r^{\prime}\,|\psi({\bf r})|^{2} V(|{\bf r}-{\bf r^{\prime}}|)|\psi({\bf r^{\prime}})|^{2}
            \ea
						
						In the case of local interactions, $V({\bf r})=\delta({\bf r})$, in which case the minimization of the energy yields the usual local Gross-Pitaevskii equation

             \ba
            \label{GP}
            &&\mu\psi=-\frac{1}{2}\nabla^{2}\psi+\frac{1}{2}(\lambda^{-2/3}{\bf r_{\perp}}^{2}+\lambda^{4/3}z^{2})\psi-4\pi k|\psi|^{2}\psi,\nonumber\\
						&&
            \ea
where $\mu$ is the Lagrange multiplier that enforces the normalization of the wavefunction.

\vskip 0.3cm
\vskip 0.3cm

{III. \bf The local case}

We need a reliable variational trial function for the nonlocally attractive condensate, if we are to reconcile it with experimental data. Therefore it must be an accurate one in the case of local interactions as well. In that case, Eq.~(\ref{GP}) holds. Far from the origin the nonlinear term is negligible, since the wavefunction is small there, essentially turning the solution of the GP equation into a Gaussian. Near the origin, however, the density of particles is quite high, resulting in a drastically different curvature of $\psi$ there. We can accommodate both of these asymptotic regions by assuming that the trial wavefunction is a superposition of two gaussians. The one with the larger exponent will be negligible at infinity, but it will contribute a lot to the curvature at the origin, while the other Gaussian will be essentially the gaussian that appears in the solution of the linearized GP equation.

Superpositions of several Gaussians have been used in the context of dipolar Bose-Einstein condensates\cite{Rau}.  It was demonstrated that the method of coupled Gaussians is a full-fledged alternative to direct numerical solutions of the Gross-Pitaevskii equation of condensates. Moreover, Gaussian wave packets are superior in that they are capable of producing both stable and unstable stationary solutions and thus of giving access to yet unexplored regions of the space of solutions of the Gross-Pitaevskii equation. As an alternative to numerical quantum simulations on multidimensional grids coupled Gaussians were used to extend the variational calculations in such a way that numerically converged results are obtained with significantly reduced computational effort compared to the exact quantum simulations, but with similar accuracy.

We adopt then the normalized trial wavefunction

\be
\label{trial}
\psi(r,z)=A(e^{-a_{1}r^{2} - b_{1}z^{2}} + t\,e^{-a_{2}r^{2}- b_{2}z^{2}})
\ee

where

\be
\label{Acoefficient}
A=\frac{2\pi^{-3/4}}{\sqrt{\frac{\sqrt{2}}{a_{1}\sqrt{b_{1}}}+\frac{8t}{(a_{1}+a_{2})\sqrt{b_{1}+b_{2}}}+\frac{t^{2}\sqrt{2}}{a_{2}\sqrt{b_{2}}}}}
\ee

We calculate now the energy, using Eq.~(\ref{E3Ddimless}) and $V({\bf r})=\delta({\bf r})$. The kinetic energy that corresponds to the gradient is

            \ba
            \label{Ekinetic}
        &&E_{k}=\frac{A^{2}\pi^{3/2}}{8}\bigl( \frac{2\sqrt{2}}{\sqrt{b_{1}}}+\frac{\sqrt{2b_{1}}}{a_{1}}+\frac{16tb_{1}b_{2}}{(a_{1}+a_{2})(b_{1}+b_{2})^{3/2}} \nonumber\\
				&&+\frac{32ta_{1}a_{2}}{(a_{1}+a_{2})^{2}\sqrt{b_{1}+b_{2}}}+\frac{\sqrt{2}(2a_{2}+b_{2})t^{2}}{a_{2}\sqrt{b_{2}}}\bigr).
            \ea
						
						The energy corresponding to the trap potential is
						
						\ba
            \label{Epotential}
        &&E_{p}=\frac{A^{2}\pi^{3/2}}{32\lambda^{2/3}}\bigl(\frac{2\sqrt{2}}{a_{1}^{2}\sqrt{b_{1}}}+\frac{32t}{(a_{1}+a_{2})^{2}\sqrt{b_{1}+b_{2}}} \nonumber\\
				&&+\frac{t\lambda^{2}\bigl(\frac{16a_{2}}{(b_{1}+b_{2})^{3/2}}+\frac{\sqrt{2}t(a_{1}+a_{2})}{b_{2}^{3/2}}\bigr)}{a_{2}(a_{1}+a_{2})} \nonumber\\
				&&+\frac{\sqrt{2}\lambda^{2}}{a_{1}b_{1}^{3/2}}+\frac{2\sqrt{2}t^{2}}{a_{2}^{2}\sqrt{b_{2}}}\bigr).
            \ea
						
						The nonlinear part of the energy is
						
						\ba
						\label{Enl}
						&&E_{nl}=-\frac{kA^{4}\pi^{5/2}}{4}\bigl(\frac{1}{a_{1}\sqrt{b_{1}}}+\frac{32t^{3}}{(a_{1}+3a_{2})\sqrt{b_{1}+3b_{2}}}\nonumber\\
						&&+\frac{t^{4}}{a_{2}\sqrt{b_{2}}}+\frac{32t}{(3a_{1}+a_{2})\sqrt{3b_{1}+b_{2}}}\nonumber\\
						&&+\frac{12\sqrt{2}t^{2}}{(a_{1}+a_{2})\sqrt{b_{1}+b_{2}}}\bigr).
						\ea
						
						Thus the total local energy is
						
						\be
						\label{Elocal}
						E_{loc}=E_{k}+E_{p}+E_{nl}.
						\ee
            
						Minimization of this energy with respect to the parameters $a_{1}$, $b_{1}$, $a_{2}$, $b_{2}$ and  $t$ will give us the ground state.

						{\bf The isotropic case}
						
						Let us examine first the case of an isotropic potential ($\lambda=1$). Then we expect $a_{1}=b_{1}$ and $a_{2}=b_{2}$. The local energy of Eq.~(\ref{Elocal}) is reduced then to a function $E_{loc}^{isotropic}$ of the parameters $a_{1}$, $a_{2}$ and $t$.
						
						We can easily find the point where the ground state becomes unstable. At that point the local isotropic energy has an extremum and the determinant of the Hessian is zero.  Thus we have 4 equations and 4 unknown parameters ($a_{1}$, $a_{2}$, $t$, $k$). The solution of these equations gives $a_{1}=0.7175$, $a_{2}=2.7665$, $t=1.1814$ and $k=0.5791$. Thus the metastability of the ground state in the local case ceases to exist at $k=0.5791$. The corresponding value obtained by Gammal et al.\cite{Gammal} was $k=0.5746$. Hence the error in the variational result is a mere 0.8\%.					
						
\begin{figure}[t]
\vskip 0.3cm
                        \includegraphics[width=0.49\textwidth]{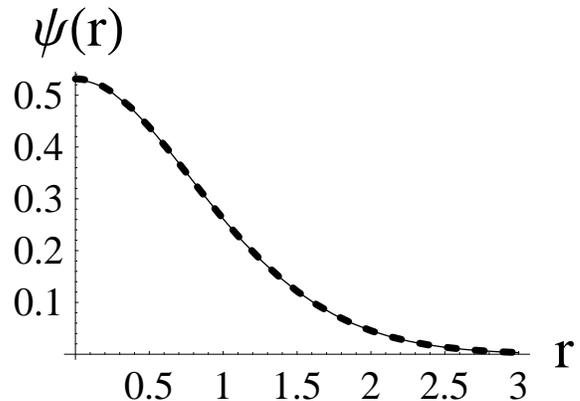}
                        \caption{\label{fig1}The numerical solution $\psi(r)$ of the isotropic GP equation for $\lambda=1$, $k=0.354039$ and $\mu=1.13224$ (dashed curve) coincides with the variational wavefunction that minimizes the energy of Eq.~(\ref{Elocal}) (continuous curve) with $k=0.354039$, $a_{1}=1.40594$, $a_{2}=0.538602$ and $t=2.76738$.}
                        \end{figure}

						We have compared the results obtained by minimizing the variational energy with the numerical solutions of the differential Eq.~(\ref{GP}), for various values of $k$, all the way from $k=0$ up to $k=0.5791$. We find that the variational solution coincides absolutely with the numerical solution. For example, for $k=0.354039$, the numerical solution of the anisotropic GP equation has $\mu=1.13224$ and an energy of 1.33405. On the other hand, the minimization of Eq.~(\ref{Elocal}) yields an energy of 1.33402, corresponding to the parameters $a_{1}=1.40594$, $a_{2}=0.538602$ and $t=2.76738$. As can be seen in Figure~\ref{fig1}, the two curves coincide.

\begin{figure}[t]
\vskip 0.3cm
                        \includegraphics[width=0.49\textwidth]{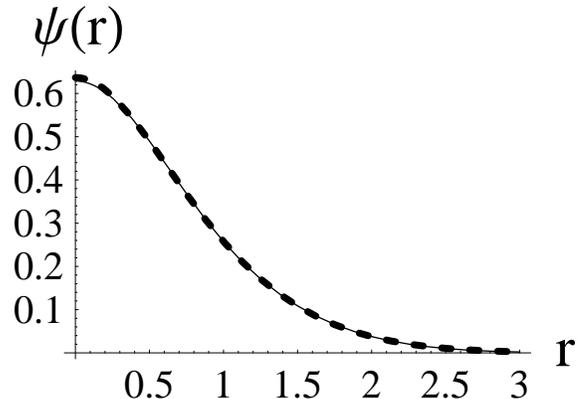}
                        \caption{\label{fig2}The numerical solution $\psi(r)$ of the isotropic GP equation for $\lambda=1$, $k=0.493367$ and $\mu=0.867908$ (dashed curve) coincides with the variational wavefunction that minimizes the energy of Eq.~(\ref{Elocal}) (continuous curve) with $k=0.493367$, $a_{1}=1.75857$, $a_{2}=0.581883$, $t=1.58232$.}
                        \end{figure}

We provide one more example for the value $k=0.493367$. The numerical solution of the GP equation ($\mu=0.867908$, energy 1.24339) coincides with the wavefunction obtained variationally (energy 1.24339, $a_{1}=1.75857$, $a_{2}=0.581883$, $t=1.58232$), as seen in Figure~\ref{fig2}. Thus we can use the superposition of two gaussians as a simple variational trial function that enables us to calculate quickly the wavefunction of the isotropic condensate.

{\bf The anisotropic case}

\begin{figure}[t]
\vskip 0.3cm
                        \includegraphics[width=0.49\textwidth]{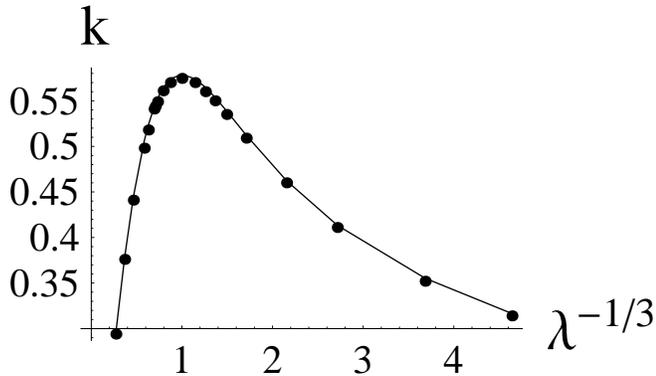}
                        \caption{\label{fig3}The critical values of $k$ at which the condensate in an anisotropic trap collapses, versus $\lambda^{-1/3}$. The dots are the results of Gammal et al.\cite{Gammal}. The continuous curve connects the corresponding results found by minimizing the variational anisotropic local energy of Eq.~(\ref{Elocal}).}
                        \end{figure}

We now examine the anisotropic case, with $\lambda=\omega_{z}/\omega_{r}$. The trial superposition of two gaussians given by Eq.~(\ref{trial}) contains five parameters. For any given value of $\lambda$ , we can find the value of $k$ at which the collapse begins by requiring that all five derivatives of the energy of Eq.~(\ref{Elocal}), as well as the determinant of the Hessian, vanish. We can then compare these variationally obtained critical values of $k$ with those obtained by Gammal et al.\cite{Gammal}. As seen in Figure~\ref{fig3}, the agreement of the variational results to those of Gammal et al. is excellent.

\begin{figure}[t]
\vskip 0.3cm
                        \includegraphics[width=0.49\textwidth]{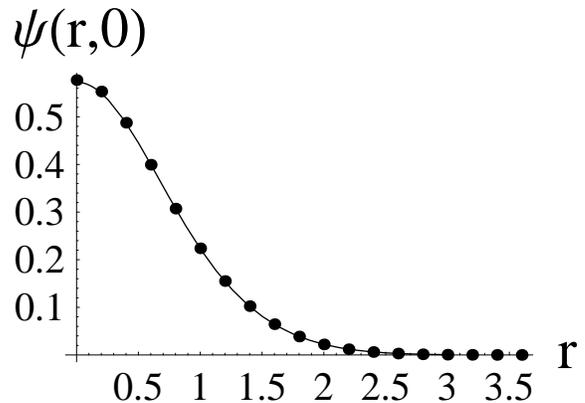}
                        \caption{\label{fig4}The variational prediction for $\psi(r,0)$ (continuous curve) coincides with the curve obtained by solving numerically the anisotropic local GP equation for the case $\lambda=6.8/17.35$ and $k=0.4$ (shown by dots).}
                        \end{figure}

\begin{figure}[t]
\vskip 0.3cm
                        \includegraphics[width=0.49\textwidth]{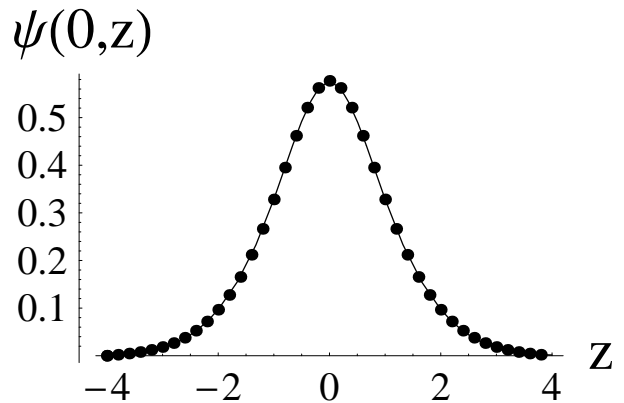}
                        \caption{\label{fig5}The variational prediction for $\psi(0,z)$ (continuous curve) coincides with the curve obtained by solving numerically the anisotropic local GP equation for the case $\lambda=6.8/17.35$ and $k=0.4$ (shown by dots).}
                        \end{figure}

We can also compare the solutions of the anisotropic GP equation obtained through elaborate numerical techniques\cite{simulations} to our variational ones. For example, if $\lambda=6.8/17.35$ and $k=0.4$, the minimization of the variational energy yields
$a_{1}=0.714249$, $b_{1}=0.347261$, $a_{2}=1.67543$, $b_{2}=1.19391$, $t=0.491734$, the corresponding energy being 1.43612. As seen in Figure~\ref{fig4} and Figure~\ref{fig5}, the variational wavefunction coincides with the one obtained by solving numerically\cite{simulations} the differential GP equation, which has the energy 1.43603.

\begin{figure}[t]
\vskip 0.3cm
                        \includegraphics[width=0.49\textwidth]{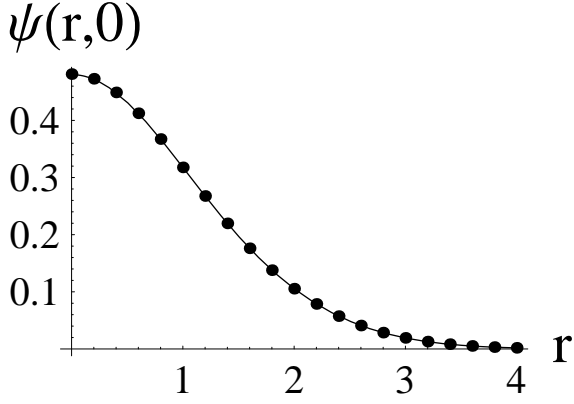}
                        \caption{\label{fig6}The variational prediction for $\psi(r,0)$ (continuous curve) coincides with the curve obtained by solving numerically the anisotropic local GP equation for the case $\lambda=4$ and $k=0.2$ (shown by dots).}
                        \end{figure}

\begin{figure}[t]
\vskip 0.3cm
                        \includegraphics[width=0.49\textwidth]{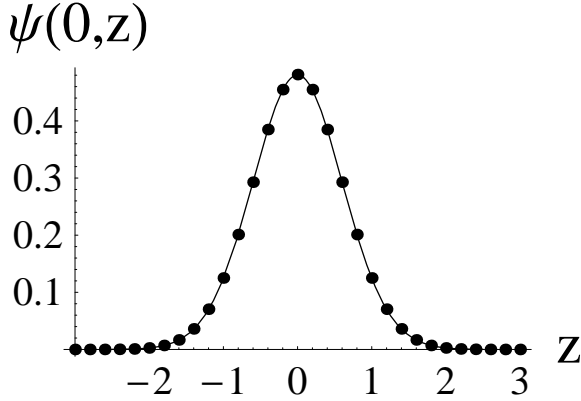}
                        \caption{\label{fig7}The variational prediction for $\psi(0,z)$ (continuous curve) coincides with the curve obtained by solving numerically the anisotropic local GP equation for the case $\lambda=4$ and $k=0.2$ (shown by dots).}
                        \end{figure}

A further example is given in in Figure~\ref{fig6} and Figure~\ref{fig7}, for the case $\lambda=4$ and $k=0.2$. The minimization of the variational energy yields $a_{1}=0.335604$, $b_{1}=1.25183$, $a_{2}=0.86238$, $b_{2}=1.97093$, $t=0.221042$, the corresponding energy being 1.802. The variational wavefunction coincides once more with the one obtained by solving numerically\cite{simulations} the differential GP equation, which has the energy 1.80199.

We conclude then that the superposition of two Gaussians is an excellent trial wavefunction for the Bose-Einstein condensate when the interactions are local, both for an isotropic and an anisotropic trap potential.

{IV. \bf Calculation of the nonlocal energy}

The Gross-Pitaevskii equation is obtained by minimizing the energy when $V({\bf r})=\delta({\bf r})$ and the interaction is local. This is just a mathematical idealization, however. The range of interactions is most certainly nozero. We shall assume then that the actual interaction is nonlocal:

\be
						\label{nonlocalinteraction}
						V({\bf r})=\frac{1}{\ell^{3}\pi^{3/2}}e^{-{\bf r}^{2}/\ell^{2}}
						\ee

In the limit $\ell\rightarrow 0$ this interaction potential becomes a delta function, since its integral ovel all space is always equal to 1. We intend to explore the size of the dimensionless parameter $\ell$, which should in any case be small. The range of the nonlocal interactions in dimensionful form will, of course, be $\ell\sqrt{\hbar/(m\omega)}$.

Our trial wavefunction will again be given by Eq.~(\ref{trial}). Then the kinetic energy and the potential energy due to the trap will still be given by Eq.~(\ref{Ekinetic}) and Eq.~(\ref{Epotential}).

The nonlinear nonlocal part of the energy can be calculated from the part of Eq.~(\ref{E3Ddimless}) that involves the nonlocal interactions and is:

\ba
\label{nonlinearnonlocal}
&&E_{nlnonloc}=-\frac{kA^{4}\pi^{5/2}}{4a_{1}\sqrt{b_{1}}(1+a_{1}\ell^{2})\sqrt{1+b_{1}\ell^{2}}}\nonumber\\
&&-\frac{8tkA^{4}\pi^{5/2}}{c_{1}\sqrt{d_{1}}}-\frac{\sqrt{2}t^{2}kA^{4}\pi^{5/2}}{c_{2}\sqrt{d_{2}}}-\frac{8t^{3}kA^{4}\pi^{5/2}}{c_{3}\sqrt{d_{3}}}\nonumber\\
&&-\frac{8t^{2}kA^{4}\pi^{5/2}}{(a_{1}+a_{2})\sqrt{b_{1}+b_{2}}(2+a_{1}\ell^{2}+a_{2}\ell^{2})\sqrt{2+b_{1}\ell^{2}+b_{2}\ell^{2}}}\nonumber\\
&&-\frac{t^{4}kA^{4}\pi^{5/2}}{4a_{2}(1+a_{2}\ell^{2})\sqrt{b_{2}(1+b_{2}\ell^{2})}}
\ea

where

\be
\label{c1}
c_{1}=a_{2}+2a_{1}^{2}\ell^{2}+3a_{1}+2a_{1}a_{2}\ell^{2},
\ee

\be
\label{c2}
c_{2}=a_{1}+a_{2}+2a_{1}a_{2}\ell^{2},
\ee

\be
\label{c3}
c_{3}=a_{1}+2a_{1}a_{2}\ell^{2}+3a_{2}+2a_{2}^{2}\ell^{2},
\ee

\be
\label{d1}
d_{1}=b_{2}+2b_{1}^{2}\ell^{2}+3b_{1}+2b_{1}b_{2}\ell^{2},
\ee

\be
\label{d2}
d_{2}=b_{1}+b_{2}+2b_{1}b_{2}\ell^{2},
\ee

\be
\label{d3}
d_{3}=b_{1}+2b_{1}b_{2}\ell^{2}+3b_{2}+2b_{2}^{2}\ell^{2}.
\ee

Thus the total nonlocal energy is
						
						\be
						\label{Enonlocal}
						E_{nonloc}=E_{k}+E_{p}+E_{nlnonloc}.
						\ee

{V. \bf A single Gaussian trial function}

We can understand qualitatively the new physics intoduced by nonlocality if we consider a single Gaussian trial function. In other words, we set $t=0$ in our trial function of Eq.~(\ref{trial}). For the sake of simplicity, we also assume an isotropic trap in this section, with $\lambda=1$ and $b_{1}=a_{1}$. Then the total nonlocal energy for the single Gaussian trial function becomes

\be
\label{epsiloniso}
\epsilon_{iso}(a_{1})=\frac{3}{8a_{1}}+\frac{3a_{1}}{2}-\frac{2ka_{1}^{3/2}}{\sqrt{\pi}(1+a_{1}\ell^{2})^{3/2}}
\ee

\begin{figure}[t]
\vskip 0.3cm
                        \includegraphics[width=0.49\textwidth]{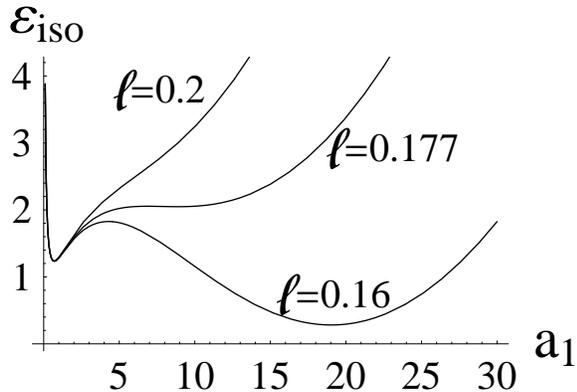}
                        \caption{\label{fig8}The energy $\epsilon_{iso}(a_{1})$ for $k=0.54758$ and for various values of $\ell$.}
                        \end{figure}

\begin{figure}[t]
\vskip 0.3cm
                        \includegraphics[width=0.49\textwidth]{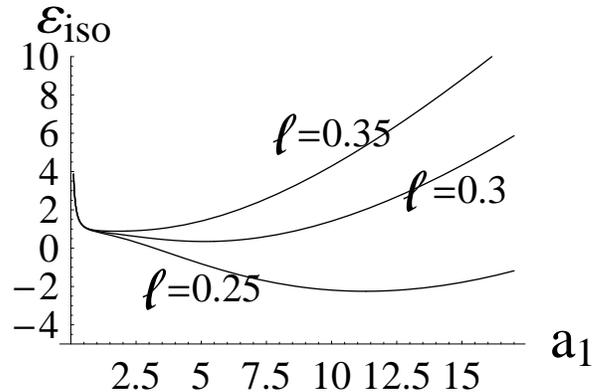}
                        \caption{\label{fig9}The energy $\epsilon_{iso}(a_{1})$ for $k=1$ and for various values of $\ell$.}
                        \end{figure}

We can understand the behaviour of $\epsilon_{iso}$ through Figure~\ref{fig8} and Figure~\ref{fig9}, for the case $k=0.54758$ and $k=1$ respectively.

For smaller values of $k$ and relatively large values of $\ell$ there is only one minimum, corresponding to a small value of $a_{1}$. However, as we decrease $\ell$, keeping $k$ fixed at a relatively small value, a second minimum appears, corresponding to a large value of $a_{1}$. As we keep decreasing $\ell$, this second minimum becomes the absolute minimum. For $k=0.54758$, for example, the abrupt transition from the minimum with small $a_{1}=0.702531$ to the minimum with large $a_{1}=15.1933$ occurs at $\ell=0.166647$. If we decrease $\ell$ further, the absolute minimum moves out towards infinite values of $a_{1}$. Thus it becomes the collapse of a condensate with local interactions, the small $a_{1}$ state corresponding to its metastable state.

For larger values of $k$ there are no local metastable solutions. However we see in Figure~\ref{fig9} that a unique nonlocal stable minimum does exist. This minimum moves out to infinite values of $a_{1}$ as $\ell\rightarrow 0$, in which case the interactions become local. Hence the collapse of the locally interacting condensate is transformed to a stable minimum for the nonlocal condensate. We note however that this stable nonlocal minimum corresponds to a narrow state with small spatial extent, since $a_{1}$ is large. It is this stable absolute minimum that corresponds to the observed remnant condensate.

{VI. \bf The observed remant condensate}

The experiments in which remnant condensates were observed(\cite{nature},\cite{thesis}) used a cylindrically symmetric cigar-shaped magnetic trap with radial and axial frequencies 17.5 Hz and 6.8 Hz, respectively. Hence $\lambda=6.8/17.5$. The critical value for the collapse of the condensate was found\cite{errorcorrected} to be $k=0.54758$.
We shall assume the trial wavefunction of Eq.~(\ref{trial}). Hence the total anisotropic nonlocal energy is given by Eq.~(\ref{Enonlocal}). Numerical minimization of this $E_{nonloc}$ shows that for $\ell=0.175201$ there are two stable minima, an anisotropic one for $a_{1}=0.756177$, $b_{1}=0.423119$, $a_{2}=1.99427$, $b_{2}=1.60734$, $t=0.795203$ with energy 1.34192, and an isotropic one for $a_{1}=19.0547$, $b_{1}=19.0345$, $a_{2}=5.74074$, $b_{2}=5.70894$, $t=0.392612$ with the same energy. Hence the observation of a collapse at $k=0.54758$ indicates that there is a phase transition at that value of $k$, between the metastable anisotropic minimum and the stable isotropic minimum. This transition requires $\ell=0.175201$. This is then the value of $\ell$ that corresponds to $k=0.54758$.

In the experiments mentioned above the scattering length $a$ is initially zero. Then it is suddenly taken to a negative value.

At the initial stage, where $a=0$, we have $k=0$. The solution of the Gross-Pitaevskii Eq.~(\ref{GP}) corresponds then to the values $t=0$, $\mu=\lambda^{-1/3}+\lambda^{2/3}/2$, $a_{1}=\lambda^{-1/3}/2$, $b_{1}=\lambda^{2/3}/2$. The corresponding initial wavefunction is just the ground state of the anisotropic harmonic oscillator:

\be
\label{psiinitial}
\phi_{0}(r,z)=\pi^{-3/4}e^{-\lambda^{-1/3}r^{2}/2-\lambda^{2/3}z^{2}/2}
\ee

If the eigenstates of the nonlinear Hamiltonian form an orthonormal set $\left\{\psi_{j}\right\}$, then the harmonic oscillator ground state $\phi_{0}$ can be written as a superposition of these nonlinear states: $\phi_{0}=\sum_{j}u_{j}\psi_{j}$, where the amplitude $u_{j}=<\psi_{j}|\phi_{0}>$. The probability of finding the system in the particular state $\psi_{f}$ is just $|u_{f}|^{2}$$=|<\psi_{f}|\phi_{0}>|^{2}$. So the fraction of atoms in the final state $\psi_{f}$ is $N_{f}/N_{0}=|<\psi_{f}|\phi_{0}>|^{2}$. Clearly, this fraction is independent of $N_{0}$, in agreement with the experimental observations.

In this case, we shall assume that the final wavefunction is the superposition of two gaussians of Eq.~(\ref{trial}). Then we find

\ba
\label{fraction}
&&\frac{N_{f}}{N_{0}}=8A^{2}\lambda^{2/3}\pi^{3/2}\big(\frac{1}{(1+2a_{1}\lambda^{1/3})\sqrt{2b_{1}+\lambda^{2/3}}}\nonumber\\
&&+\frac{t}{(1+2a_{2}\lambda^{1/3})\sqrt{2b_{2}+\lambda^{2/3}}}\big)^{2}
\ea

The experimental procedure involves a sudden change of the scattering length from $a=0$ to $a=a_{f}$, and therefore from $k=0$ to $k=|a_{f}|N_{0}/d=k_{f}$. Close to the Feschbach resonance the self-interactions of the condensate are controlled by the magnetic field. Clearly, this will not hold just for the scattering length $a$, but also for the interaction length $\ell$. In other words, when $a$ changes to a value $a_{f}$, we expect $\ell$ to change also to a value $\ell_{f}$.

\begin{figure}[t]
\vskip 0.3cm
                        \includegraphics[width=0.49\textwidth]{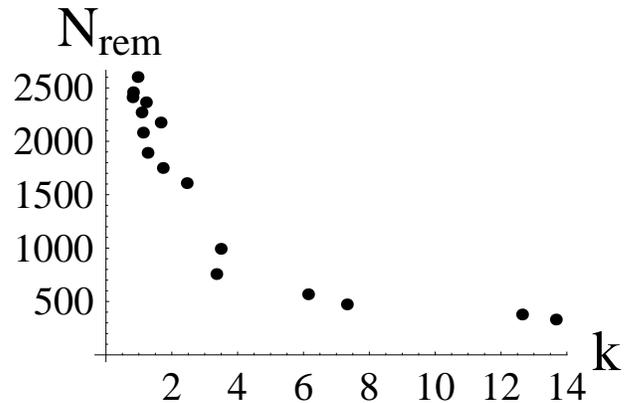}
                        \caption{\label{fig10}The observed number of atoms in the remnant condensate $N_{rem}$ as a function of the final value $k_{f}$ of $k$, from the data of figure 4.7 of Ref. \cite{thesis}, with zero initial scattering length and $N_{0}=6000$.}
                        \end{figure}

We shall try to explain the data of figure 4.7 of Ref. \cite{thesis}, shown in Figure~\ref{fig10} in the form of the number of atoms in the remnant condensate $N_{rem}$ versus the final value of $k$, $k_{f}=|a_{f}|N_{0}/d$. For these data the number of atoms in the initial condensate and the initial value of $a$ were $N_{0}=6000$ and $a=0$, respectively. The value of $d$ was 3.05$\mu m$.

\begin{figure}[t]
\vskip 0.3cm
                        \includegraphics[width=0.49\textwidth]{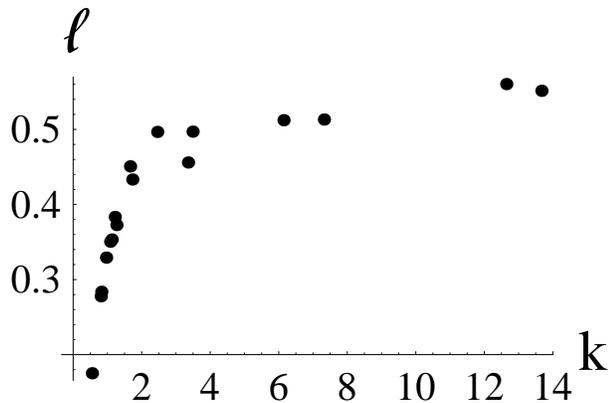}
                        \caption{\label{fig11}The value of the range $\ell$ of the nonlocal interaction as a function of $k$. The values were derived from the trial double gaussian wavefunction using the experimental data of figure 4.7 of Ref. \cite{thesis}, with zero initial scattering length and $N_{0}=6000$.}
                        \end{figure}

We first attempted to fit these data to a single gaussian trial function ($t=0$). We found out that there were no minima of the energy that led to the observed values of the fraction $N_{f}/N_{0}$. We then tried the superposition of two gaussians. For each point in the experimental figure we sought the values of the five trial parameters ($a_{1}$, $b_{1}$, $a_{2}$, $b_{2}$, $t$) and the value of $\ell$ that would render all five derivatives of the energy equal to zero and that would give the observed value of $N_{f}/N_{0}$. We would then verify that these values did in fact correspond to a minimum of the nonlocal energy for that particular value of $\ell$. For example, the observed remnant fraction for $k_{f}=0.807$ was 0.402. This fraction is the one obtained at the minimum of the energy when $\ell=0.277803$, in which case we have $a_{1}=1.65984$, $b_{1}=1.54306$, $a_{2}=5.63176$, $b_{2}=5.55371$, $t=2.23696$. We use in this way all the points in the experimental figure in order to derive the values of $\ell$ that correspond to each final value of $k$. We also include the separate point that was derived above, i.e. that the value $\ell=0.175201$ corresponds to $k=0.54758$. We plot the value of $\ell$ versus the corresponding value of $k$ in Figure~\ref{fig11}.

We shall now derive an expression for $\ell$ as a function of $k$. We expect both $\ell$ and $a$ to be functions of the magnetic field near the Feshbach resonance. Indeed, in that region $a$ should be a linear function of $1/(B-B_{0})$, where $B_{0}$ is the resonance position\cite{errorcorrected}. We have thus

\be
\label{fitell}
a=-\frac{s_{1}d}{N_{0}}-\frac{s_{2}d}{N_{0}}\frac{B_{0}}{B-B_{0}},
\ee

with $s_{1}$ and $s_{2}$ unknown parameters. Hence

\be
\label{fitell2}
B=B_{0}\frac{k-s_{1}+s_{2}}{k-s_{1}}
\ee

Furthermore we expect on dimensional grounds $\ell^{2}$ to be proportional to $1/B$, i.e. $\ell=\sqrt{qB_{0}/B}$. Consequently

\be
\label{formulaell}
\ell=\frac{\sqrt{q}\sqrt{k-s_{1}}}{\sqrt{k-s_{1}+s_{2}}},
\ee

where $s_{1}$, $s_{2}$ and $q$ are parameters to be determined by fitting this expression to the points of Figure~\ref{fig11}.

\begin{figure}[t]
\vskip 0.3cm
                        \includegraphics[width=0.49\textwidth]{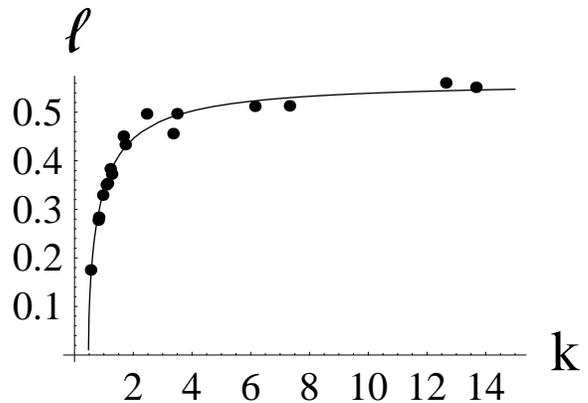}
                        \caption{\label{fig12}The value of the range $\ell$ of the nonlocal interaction as a function of $k$. The dots are values that were derived from the trial double gaussian wavefunction using the experimental data of figure 4.7 of Ref. \cite{thesis}, with zero initial scattering length and $N_{0}=6000$. The continuous curve corresponds to the expression of Eq.~(\ref{formulaell}) for $s_{1}=0.475682$, $s_{2}=0.920563$, $q=0.318639$.}
                        \end{figure}

Indeed, this fitting yields $s_{1}=0.475682$, $s_{2}=0.920563$, $q=0.318639$. The expression of Eq.~(\ref{formulaell}) fits quite well the values of $\ell$ which were derived using the double Gaussian trial wavefunction, as we can see in Figure~\ref{fig12}.

{\bf Conclusions}

We have shown that a nonlocal extension of the Gross-Pitaevskii equation can explain the stability of the remnant condensate that was observed during the collapse of attractive Bose-Einstein condensates. The reason for this stability is the existence of an absolute stable minimum that appears only if the energy is nonlocal. As the scattering length is tuned from zero scattering length to a negative scattering length, the condensate makes a sudden transition to the absolute minimum that corresponds to the new scattering length. The fraction of the number of atoms that remains in the remnant condensate is simply the square of the modulus of the overlap integral between the initial and final states. Consequently, one needs to know as well as possible the final wavefunction. We have shown that a superposition of two Gaussians is an excellent approximation to the solution of the usual anisotropic GP equation for an attractive condensate. We have used this trial function in order to find the minimum of the final nonlocal energy and the corresponding overlap integral. The magnetic field determines the scattering length, as well as the behaviour of the range $\ell$ of the nonlocal interactions. We have used the superposition of two Gaussians in order to find the values of the range $\ell$ of nonlocal interactions that correspond to the observed remnant fractions of atoms, as a function of the final scattering length. A simple expression derived for $\ell$ near the Feshbach resonance seems to represent well the experimental data.

\end{document}